# Blueprint for deterministic all-optical switching of magnetization

C. S. Davies[1,2,*], T. Janssen[1,2], J. H. Mentink[2], A. Tsukamoto[3], A. V. Kimel[2], A. F. G. van der Meer[1,2], A. Stupakiewicz[4], and A. Kirilyuk[1,2,†]

[1] *FELIX Laboratory, Radboud University, 7c Toernooiveld, 6525 ED Nijmegen, The Netherlands*

[2] *Radboud University, Institute for Molecules and Materials, 135 Heyendaalseweg, 6525 AJ Nijmegen, The Netherlands*

[3] *College of Science and Technology, Nihon University, 7-24-1 Funabashi, Chiba 274-8501, Japan*

[4] *Laboratory of Magnetism, Faculty of Physics, University of Bialystok, 1L Ciolkowskiego, 15-245 Bialystok, Poland*

## Abstract

We resolve a significant controversy about how to understand and engineer single-shot all-optical switching of magnetization in ferrimagnets using femto- or picosecond-long heat pulses. By realistically modelling a generic ferrimagnet as two coupled macrospins, we comprehensively show that the net magnetization can be reversed via different pathways, using a heat pulse with duration spanning all relevant timescales within the non-adiabatic limit. This conceptual understanding is fully validated by experiments studying the material and optical limits at which the switching process in GdFeCo alloys loses its reliability. Our interpretation and results constitute a blueprint for understanding how deterministic all-optical switching can be achieved in alternative ferrimagnets using short thermal pulses.

[*] Corresponding author: C.Davies@science.ru.nl
[†] Corresponding author: A.Kirilyuk@science.ru.nl

The societal thirst for smaller, faster and more energy-efficient hard-disk drive technology stimulates intense research devoted to finding and understanding magnetization switching processes. The industrially-favored approach uses just a writing magnetic field, but the superparamagnetic effect[1] and the associated recording trilemma impedes further improvements in this direction. In 2007, the discovery of all-optical switching (AOS)[2], in which ultrashort optical pulses reverse magnetization without assisting magnetic fields, gave birth to the identification and study of a whole family of AOS-related effects[3,4,5] displayed by different materials and enabled by tailored optical pulses. To date, however, the only materials that have been known to display ultrafast single-shot AOS are amorphous alloys of GdFeCo[3] and multilayered stacks of Pt/Gd/Co[6]. In these materials, a single optical pulse will toggle the magnetization deterministically i.e. irrespective of its initial polarity.

Since the first discovery of AOS in GdFeCo, many reports have aimed to elucidate its underlying mechanisms, but due to the undeniable complexity of ultrafast magnetism[7], many conflicting results and interpretations have emerged. It was initially thought that the inverse Faraday Effect drove helicity-dependent AOS in GdFeCo[3], but later careful and quantitative analysis of the wavelength-dependent fluence requirements irrefutably revealed that magnetic circular dichroism was responsible[8], in combination with deterministic AOS. It was also assumed[3,9-11] that a sub-picosecond optical pulse was a compulsory prerequisite for deterministic AOS, but experimental reports have shown that even 15-picosecond-long pulses can suffice in certain cases[12,13].

While it is clear that a femtosecond-long laser pulse generates a strongly non-equilibrium state in GdFeCo with a fully-demagnetized FeCo-sublattice[3,9,10-11], it is also clear that laser-pulses with duration $\tau$ longer than the electron-lattice interaction $\tau_{e-l}$ (~2 ps) cannot induce dramatic overheating of the free electrons[12,13]. Such overheating is crucial for ultrafast demagnetization as the spin-lattice relaxation rate is proportional to the effective electron temperature[14,15]. The community of ultrafast magnetism therefore found it counter-intuitive that deterministic AOS could be achieved using laser and current pulses with $\tau > 10$ ps. These experimental observations even led to statements[13] about the insolvency of the mechanism via a strongly non-equilibrium state. Such statements, however, overlook the fact that the three-temperature model[7,12-13,16] does not always adequately represent a ferrimagnet. Two very different magnetic sublattices are better represented not by one, but by two interconnected reservoirs, where the characteristic time of interaction $\tau_{Gd-FeCo}$ between the spin-reservoirs of Gd and FeCo is defined by the inter-sublattice exchange interaction.

Because of this, fast change of the magnetization of one of the sublattices is possible pie at the cost of the other another, and does not require a spike in the electronic temperature. The overarching criterion for deterministic AOS lies in the condition that the heating induces a strongly non-equilibrium state. If this is satisfied, even relatively slow heating of the system triggering purely exchange-driven dynamics can achieve reversal, provided that (i) there is more angular momentum in the Gd sublattice than in the FeCo one, and (ii) the spin-lattice thermalization time is slower than $\tau_{Gd-FeCo}$. This leads to the observable transient ferromagnetic state, whereby the magnetization of FeCo crosses zero while Gd is still demagnetizing, which is a compulsory prerequisite for deterministic AOS.

In this letter, we present a conceptual understanding of deterministic AOS derived for a generic ferrimagnet of composition $A_{100-x}B_x$, using laser pulses with duration covering all relevant time scales. The magnetization dynamics of $AB$, which underpin the switching process, can be described using a master/slave relationship, with $A$ being the "master" and $B$ serving as the "slave". Two distinct pathways allow for deterministic AOS, either with angular momentum flowing from both sublattices to the external environment or between $A$ and $B$ themselves. The direction of the flow is dictated by the combination of the relative concentrations of $A$ and $B$ and the temporal properties of the excitation. To validate our conceptual understanding, we use a phenomenological mean field theory describing the sublattice-resolved longitudinal magnetization dynamics of $A_{100-x}B_x$, taking in to account both the temporal profile of a thermal load and the alloy composition. To provide ultimate proof of our interpretation, we experimentally study the material and optical parameters that enable or disable deterministic AOS in $Gd_x(FeCo)_{100-x}$ alloys. Specifically, we identify a critical pulse-duration threshold that defines the deterministic character of AOS, and increases monotonically with the concentration $x$ of slave gadolinium. Photons in a very wide spectral range, from the visible to mid-infrared, are also shown to be equally capable of triggering deterministic AOS. Our conceptual interpretation explains both our measurements and a wealth of other experimental and numerical findings that have, until now, not been unified within a common framework of understanding. Moreover, we believe our understanding may be expanded to experimentally predict the general conditions that will enable deterministic AOS in different materials.

The master/slave relationship intrinsic to our considered generic ferrimagnet $AB$ derives from the fact that, in isolation, sublattices $A$ and $B$ are ferromagnetic and paramagnetic respectively. Nevertheless, the intersublattice exchange coupling gives rise to a common Curie temperature in equilibrium, and also the existence of two degenerate equilibrium states, with *A* and *B* having antiparallel magnetization. These two states are indicated by green dots in the sublattice-resolved phase diagram of angular momentum *S* shown in Fig. 1, and trajectories connecting the two correspond to deterministic AOS pathways[17]. Under equilibrium conditions, it is impossible for AOS to occur without an assistive magnetic field. Adiabatic heating of the ferrimagnet, i.e. $\tau > \tau_{s-l}$ where $\tau_{s-l}$ is the spin-lattice thermalization time, results in $S_B$ decreasing more rapidly than $S_A$ (inset of Fig. 1), and ends with the complete destruction of magnetization. This scenario corresponds to the dashed trajectory shown in Fig. 1.

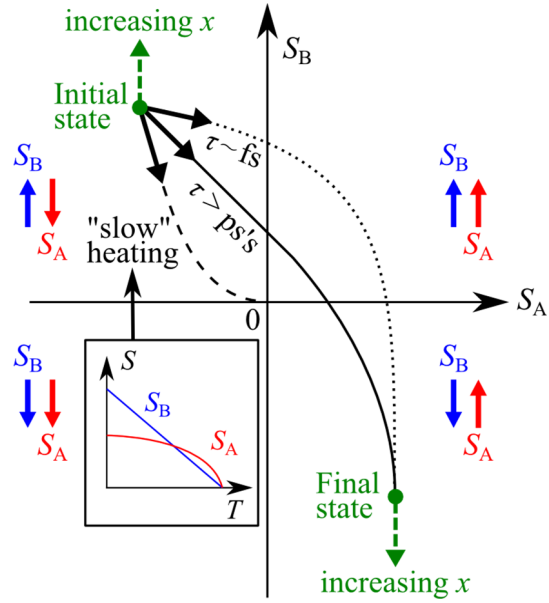

**Fig. 1** Conceptual phase map showing the different pathways for thermally-induced relaxation and recovery of the constituent-resolved angular momentum *S* of the ferrimagnet $A_{100-x}B_x$. The thick green dots indicate positions of equilibrium, and by varying *x*, these states are translated across the map. Excitation of the ferrimagnet by thermal pulses of varying duration $\tau$ lead to different trajectories. Shown in the inset is the adiabatic thermal dependence of the angular momentum.

When the ferrimagnet is instead heated under non-equilibrium conditions, the magnetization can relax via two distinct mechanisms. The first involves inter-sublattice exchange coupling (with a characteristic timescale $\tau_{A-B}$) whereby the angular momentum of the master sublattice grows at the expense of the slave's. If the dynamics are driven purely by exchange coupling, the total angular momentum of *AB* is conserved and so $\partial_t S_A = -\partial_t S_B$. As one therefore reduces $\tau$ from above to below $\tau_{s-l}$, the solid trajectory shown in Fig. 1 becomes increasingly linear along the figure diagonal (solid curve). Provided that (i) there is more angular momentum in slave-*B* than in master-*A*, and (ii) $\tau_{A-B} < \tau_{s-l}$, relatively slow heating of the system (> 10 ps) can still satisfy the observable condition for deterministic AOS (that $S_A$ crosses zero while $S_B$ is demagnetizing). Upon forming the transient ferromagnetic state, continuous exchange of angular momentum leads to the slave switching its magnetization polarity, as dictated by master *A*, and so deterministic AOS is successfully achieved.

Upon further reduction of $\tau$ towards the timescale of electron-lattice thermalization (~2 ps in GdFeCo)[12,16], temperature-induced dissipation of $S_A$ and $S_B$ to the external environment overwhelms the exchange coupling, and the sublattices essentially relax independently. Furthermore, if *A* has a smaller spin than *B*, *A* will demagnetize faster[9], resulting in a reasonably-horizontal dotted trajectory as indicated in Fig. 1 (in GdFeCo, this gradient is approximately 4:1)[9]. $S_B$ is now even larger when $S_A$ crosses zero, and so the already-cooling system enables the intersublattice exchange coupling and subsequent magnetization recovery to complete the switching process.

By varying the alloy concentration of $A_{100-x}B_x$, the initial and final equilibrium states (green dots in Fig. 1) will shift. Increasing *y* shifts the initial and final equilibrium states of *AB* up and down respectively in Fig 1, allowing a steeper trajectory to join the two states. Physically, the slave has more angular momentum available to transfer to the master, enabling a longer pulse (still satisfying the non-adiabatic condition $\tau_{A-B} < \tau_{s-l}$) achieve deterministic AOS. Conversely, reducing *y* will disable the possibility for deterministic AOS to proceed via exchange coupling only, if $|S_B| < |S_A|$. However, a shorter pulse generating a more horizontal trajectory in Fig. 1 would still suffice.

To numerically test our conceptual understanding summarized in Fig. 1, we have expanded upon the phenomenological mean-field model of relaxation dynamics of a ferrimagnet developed by Mentink *et al*[10,18]. In this model, the longitudinal dynamics of the $S_A$ and $S_B$ is governed by the interplay between the inter-sublattice exchange and spin-lattice relaxation of individual sublattices. The coupled equations of motion characterizing the temperature-dependent angular momentum of each sublattice (which are treated as a pair of macrospins) are

$$\frac{dS_A}{dt} = -\lambda_e(H_B - H_A) + \lambda_A H_A, \quad (1)$$

$$\frac{dS_B}{dt} = \lambda_e(H_B - H_A) + \lambda_B H_B, \quad (2)$$

where $\lambda_A$ and $\lambda_B$ characterize the flow of angular momentum from the indicated sublattice to the external environment (of temperature $T$), $\lambda_e$ characterizes the inter-sublattice exchange, and $H$ represents the effective field acting on the subscripted sublattice. A full description of the mean-field model is supplied[19] in Supplemental Note 1.

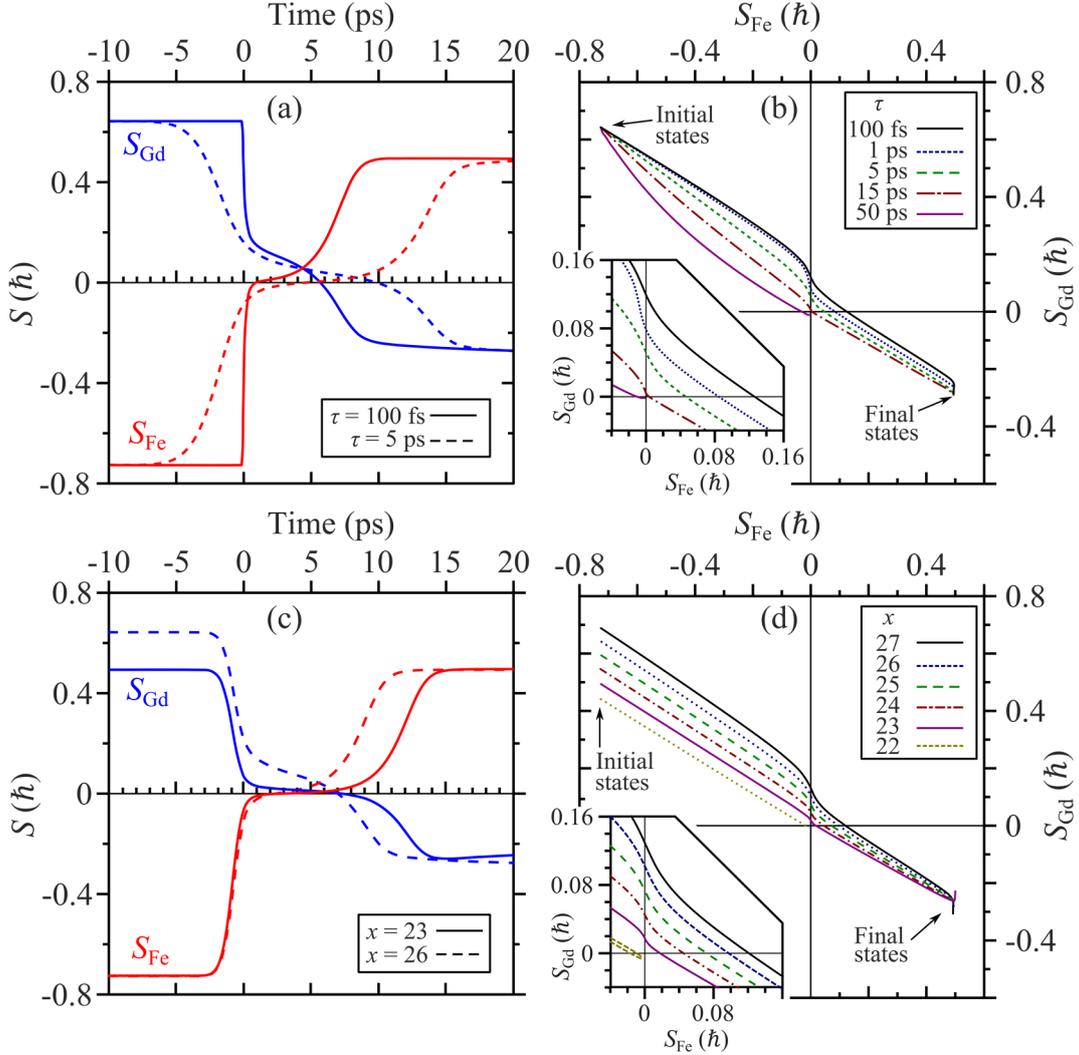

**Fig. 2**  (a) Calculated time-resolved dynamics of the angular momentum $S$ of master Fe (red line) and slave Gd (blue line) in the ferrimagnet $Gd_{26}Fe_{74}$ triggered using different pulse durations $\tau$ as indicated. (b) Corresponding phase map of the sublattice-resolved angular momentum trajectories of the ferrimagnet $Gd_{26}Fe_{74}$ obtained using different $\tau$. Shown in the inset is a zoomed section. (c)-(d) Same as in panels (a)-(b) except $\tau = 2$ ps and the alloy concentration of the ferrimagnet $Gd_xFe_{100-x}$ is varied.

Using Eqs. (1)-(2), we calculated the sublattice-resolved magnetization dynamics of $AB$ with different alloy concentrations in response to thermal pulses of varying duration. Material parameters typical of the ferrimagnetic alloy $Gd_x(FeCo)_{100-x}$ were adopted, taking the transition metal component as a single sublattice and using concentration-independent material properties (thus restricting the independent parameters to just $\lambda_e$, $\lambda_A$ and $\lambda_B$). The full-width half-duration $\tau$ of the temporally-Gaussian pulse enters the model through a time-dependent temperature that captures the spirit of the two-temperature model.

Figure 2 (a) shows the results of the calculations for the alloy $Gd_{26}Fe_{74}$, obtained with varying pulse duration $\tau$. With $\tau = 100$ fs (solid curves), we successfully achieve deterministic AOS via different demagnetization rates and the clear formation of a transient ferromagnetic state. Generally, increasing $\tau$ leads to increasingly comparable demagnetizing rates of Gd and Fe. Stretching the pulse duration to 5 ps (dashed curves) still enables deterministic AOS, but $S_{Gd}$ and $S_{Fe}$ almost completely quench simultaneously. In practice, thermal fluctuations may dominate at this point, and the switching would lose its deterministic character. Upon stretching the pulse duration even further (Supplemental Note 2)[20], the polarity of the transient ferromagnetic state undergoes reversal[21] i.e. the magnetization of the slave switches before that of the master. This is consistent with both the experimental and numerical results reported in Refs. [21]-[23], and we observe in this case that AOS always fails.

By recasting the time-resolved trajectories of $S_{Gd}$ and $S_{Fe}$ as functions of each other, we gain a numerically-supported insight of how the pulse duration controls the process of deterministic AOS. Figure 2 (b) shows[20] that by increasing $\tau$, the AOS trajectory initially becomes more linear, and then curves below the figure diagonal, reflecting the increasing dominance of the inter-sublattice exchange coupling. By repeating the same calculations for $Gd_xFe_{100-x}$ alloys with varying $x$ and fixed pulse duration $\tau = 2$ ps (Fig. 2 (c)), the initial ferrimagnetic state in the plane $S_{Fe}$-$S_{Gd}$ is shifted upwards (Fig. 2 (d)). This permits a steeper gradient of the AOS trajectory where $S_{Fe}$ crosses zero while $S_{Gd}$ is still demagnetizing. physically allowing a ferrimagnet with more slave constituents to be deterministically switched using a longer pulse.

To obtain ultimate experimental evidence of our interpretation, we performed a set of experiments exposing 6 GdFeCo alloys with different sublattice concentrations to single laser pulses of varying duration. The samples were all of elemental composition $Gd_x(FeCo)_{100-x}$, with $22 \leq x \leq 27$, and all possessed out-of-plane magneto-crystalline anisotropy. Specific details of the samples are supplied[24] in Supplemental Note 3. The laser pulses had a photon energy of 1.55 eV (central wavelength 800 nm) and a duration that could be adjusted between 60 fs and 6.0 ps and resolved with an accuracy of < 100 fs. The effect of the optical pulse on the sample magnetization at room temperature was monitored using a magneto-optical microscope sensitive to the out-of-plane component of magnetization via the Faraday effect.

The insets of Fig. 3 show typical magneto-optical images recorded for the alloy $Gd_{23}(FeCo)_{77}$ after exposure to a single laser pulse of duration $\tau = 1.4$ ps (bottom-right inset) and $\tau = 1.5$ ps (top-left inset). Deterministic AOS is clearly observed in the former, whereas the latter displays a random spatial distribution of magnetic domains i.e. demagnetization. Further measurements showed that pulse durations below and above 1.4 ps always result in deterministic AOS and demagnetization respectively, and thus we conclude that $Gd_{23}(FeCo)_{77}$ possesses a critical threshold $\tau_c = 1.4$ ps whereby deterministic AOS is enabled if $\tau < \tau_c$ but is disabled if $\tau > \tau_c$.

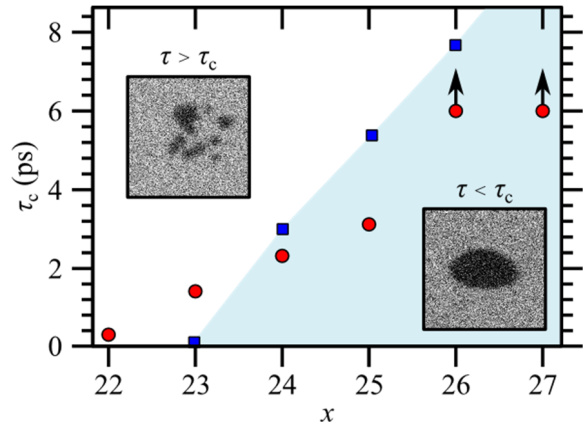

**Fig. 3** The critical pulse-duration threshold $\tau_c$ is plotted (red circles) as a function of alloy composition for $Gd_x(FeCo)_{100-x}$, measured using pulses of photon energy 1.55 eV. Deterministic AOS is achieved if $\tau < \tau_c$, but disabled if $\tau > \tau_c$. Experimentally we could only realize $\tau \leq 6$ ps, and so can only conclude that $\tau_c > 6$ ps for $x \geq 26$. Also shown are the calculated values of $\tau_c$ for the alloy $Gd_xFe_{100-x}$ (blue squares). Insets: Typical background-corrected magneto-optical images, of side length 100 μm, obtained for $Gd_{23}(FeCo)_{77}$ showing deterministic AOS (bottom-right panel, $\tau = 1.4$ ps) and demagnetization (top-left inset, $\tau = 1.5$ ps). The contrast in the images is proportional to the out-of-plane component of magnetization.

We repeated the measurements shown in the insets of Fig. 3 for each $Gd_x(FeCo)_{100-x}$ alloy, and presented in Fig. 3 are the corresponding thresholds $\tau_c$ as a function of $x$. Clearly, as the percentage of the slave gadolinium in $Gd_x(FeCo)_{100-x}$ increases, the pulse duration still capable of enabling deterministic AOS increases monotonically. When $x \geq 26$, we were unable to identify the threshold which exceeded 6.0 ps (a limit imposed by our regenerative amplifier). However, in Ref. [13], $\tau_c = 15$ ps for $x = 27.5$, which is in good agreement with the implications of our results. Using the calculations, we obtain the same linear trend, taking in to account that thermal fluctuations disable deterministic AOS if $S_{Fe}$ and $S_{Gd}$ cross zero almost simultaneously[25]. These findings are clearly in excellent agreement with our conceptual understanding, demonstrating the deep physical insight one can obtain by considering AOS trajectories across the $S_A$-$S_B$ plane.

A fundamental assumption of our model lies in our use of the concept of "temperature". Temperature can be associated with equilibrium phenomena only[26], but it is routinely used in descriptions of non-equilibrium magnetization dynamics[3,9,10-11,16]. An optical excitation of high photon energy 1.55 eV stimulates a multitude of intra- and inter-band electronic excitations, causing the temperature of the spins to become poorly defined[13]. The importance of these high-energy excitations in the effectiveness of the demagnetization process was also a subject of recent theoretical debate[27-28]. As an efficient and fast demagnetization is an essential prerogative for switching in our model, we can provide a direct experimental answer to this problem by considerably reducing the photon energy of the optical excitation. We therefore use pulses in the mid-infrared spectral range at FELIX (Free Electron Lasers for Infrared eXperiments)[29-30]. A single optical pulse, with photon energy ranging between $E = 70$ meV and $E = 230$ meV, is focused to a spot of diameter 100 μm[31] on the surface of the GdFeCo samples. The duration of the pulse is controlled through cavity desynchronization[32], allowing the latter to be varied between ~400 fs and at least ~6.5 ps[33].

Figure 4 shows the experimentally-measured state map for $Gd_x(FeCo)_{100-x}$ with $x = 24$, while the state maps for $x = 25$ and $x = 26$ are provided[34] in Supplemental Note 6. In these maps, we summarize how the deterministic character of AOS in $Gd_x(FeCo)_{100-x}$ alloys depends on the photon energy and pulse-duration, obtained through analyzing magneto-optical images recorded after exposing the material to consecutive optical pulses[34]. For all the studied compositions of $Gd_x(FeCo)_{100-x}$, we generally observed that the photon energy, despite being adjusted by a factor of more than 20 (between 70 meV and 1.55 eV), always enabled deterministic AOS provided the pulse duration was sufficiently low. This result validates both the microscopic picture of ultrafast demagnetization advanced by Schellekens *et al* in Ref. [28] and the invocation of temperature in our model. Moreover, these results confirm that relatively gentle heating of the free electrons in GdFeCo is sufficient to achieve the necessary strongly non-equilibrium state required for deterministic AOS.

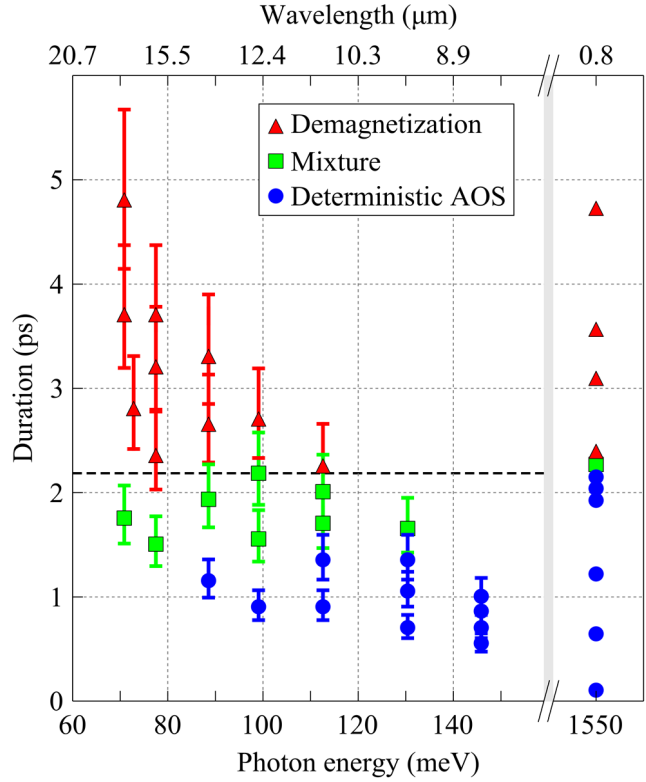

**Fig. 4**  State map recorded for $Gd_{24}(FeCo)_{76}$ indicating how the switching process depends on the photon energy and pulse duration. Points indicated with a blue circle or red triangle correspond to observations of deterministic AOS or demagnetization respectively, whereas green triangles correspond to observations of both effects arising from jitter in the pulse duration.

In summary, we have revealed a new conceptual understanding of the mechanism underpinning deterministic AOS. We base our description on there being a master/slave relationship between the constituents of a generic ferrimagnet *AB*, where *A* (the master) is ferromagnetic and *B* (the slave) is paramagnetic in isolation. Deterministic AOS can be achieved through two distinct pathways, either by angular momentum flowing from *A* and *B* to the external bath or through angular momentum being transferred from the slave to the master. The choice of which pathway is followed depends solely on the pulse duration relative to the timescales of the spin-lattice and inter-sublattice exchange interactions, and increasing the concentration of slaves in *AB* also increases the pulse duration that can still enable deterministic AOS. We use a phenomenological mean field approach to validate our understanding, and provide ultimate proof by studying how the critical pulse-duration threshold (above/below which deterministic AOS is disabled/enabled) evolves as a concentration of the slave in GdFeCo alloys. Moreover, by demonstrating that mid-infrared optical pulses are capable of realizing deterministic AOS, we experimentally show that the three-temperature model offers a valid description of magnetization dynamics, provided that suitable discrimination is made between the spin-reservoirs of *A* and *B*. We believe our conceptual understanding resolves many controversies surrounding deterministic AOS, and could be deployed to understand how deterministic AOS can be achieved in a larger class of materials.

**Acknowledgements**

The authors thank S. Semin, T. Toonen and all technical staff at FELIX for technical support. This research has received funding from the European Union's Horizon 2020 research and innovation program under FET-Open Grant Agreement No. 713481 (SPICE), de Nederlandse Organisatie voor Wetenschappelijk Onderzoek (NWO), the project TEAM/2017-4/40 of the foundation for Polish Science, and the Grant-in-Aid for Scientific Research on Innovative Area, "Nano Spin Conversion Science" (Grant No. 26103004).

[32] R. J. Bakker, D. A. Jaroszynski, A. F. G. van der Meer, D. Oepts and P. W. van Amersfoort, "Short-pulse effects in a free-electron laser." *IEEE J. Quantum Electron.* **30**, 1635 (1994).

[33] See Supplemental Note 5 at [url inserted by publisher] for a brief description of how we calculate the duration of the mid infra-red pulses.

[34] See Supplemental Note 6 at [url inserted by publisher] for the state maps measured for the GdFeCo samples with different concentrations of gadolinium, and exemplary images showing the three processes that are triggered by the pulses (deterministic AOS, demagnetization, and a mixture of the latter two).